\pdfoutput=1
\documentclass[11pt,twoside,a4paper,cmspaper,final,collab]{cms-tdr}
\def\svnVersion{28f412d}\def\svnDate{2024/02/10}\def\cmsCernNoTag{CERN-EP-2023-234}\def\cmsCernDate{\today}\def\cmsMessage{Published in Physics Letters B as \href{http://dx.doi.org/10.1016/j.physletb.2024.138478}{\doi{10.1016/j.physletb.2024.138478}.}}
\begin{document}\cmsNoteHeader{TOP-22-010}

 \ifthenelse{\boolean{cms@external}}{\providecommand{\cmsLeft}{upper\xspace}}{\providecommand{\cmsLeft}{left\xspace}} 
 \ifthenelse{\boolean{cms@external}}{\providecommand{\cmsRight}{lower\xspace}}{\providecommand{\cmsRight}{right\xspace}} 
\newcommand{\PA}{\HepParticle{A}{}{}\Xspace}
\newcommand{\mA}{\ensuremath{m_{\PA}}\xspace}
\newcommand{\mH}{\ensuremath{m_{\PH}}\xspace}
\newcommand{\mh}{\ensuremath{m_{\Ph}}\xspace}
\newcommand{\ttc}{\ensuremath{\PQt{}\PQt{}\PAQc}\xspace}
\newcommand{\ttu}{\ensuremath{\PQt{}\PQt{}\PAQu}\xspace}
\newcommand{\ttq}{\ensuremath{\PQt{}\PQt{}\PAQq}\xspace}
\newcommand{\zz}{\ensuremath{\PZ\PZ}\xspace}
\newcommand{\wz}{\ensuremath{\PW\PZ}\xspace}
\newcommand{\ttW}{\ensuremath{\ttbar\PW}\xspace}
\newcommand{\ttZ}{\ensuremath{\ttbar\PZ}\xspace}
\newcommand{\ttH}{\ensuremath{\ttbar\PH}\xspace}
\newcommand{\ttV}{\ensuremath{\ttbar\PV}\xspace}
\newcommand{\zjet}{\ensuremath{\PZ}+\text{jets}\xspace}
\newcommand{\wjet}{\ensuremath{\PW}+\text{jets}\xspace}
\newcommand{\vjet}{\ensuremath{\PV}{}+\text{jets}\xspace}
\newcommand{\ttVH}{\ensuremath{\ttbar\PV\PH}\xspace}
\newcommand{\ttVV}{\ensuremath{\ttbar\PV\PV}\xspace}
\newcommand{\tZ}{\ensuremath{\PQt\PZ}\xspace}
\newcommand{\tW}{\ensuremath{\PQt\PW}\xspace}
\newcommand{\tbarW}{\ensuremath{\PAQt\PW}\xspace}
\newcommand{\tttj}{\ensuremath{\ttbar\PQt\text{j}}\xspace}
\newcommand{\tttW}{\ensuremath{\ttbar\PQt\PW}\xspace}
\newcommand{\tttt}{\ensuremath{\ttbar\ttbar}\xspace}
\newcommand{\tZq}{\ensuremath{\PQt\PZ\PQq}\xspace}
\newcommand{\VVV}{\ensuremath{\PV\PV\PV}\xspace}
\newcommand{\VV}{\ensuremath{\PV\PV}\xspace}
\newcommand{\WLLJJ}{\ensuremath{\PW\Pell\Pell\PQq\PQq}\xspace}
\newcommand{\ee}{\ensuremath{\Pepm\Pepm}\xspace}
\newcommand{\mumu}{\ensuremath{\PGmpm\PGmpm}\xspace}
\newcommand{\emu}{\ensuremath{\Pepm\PGmpm}\xspace}
\newcommand{\pp}{\ensuremath{\Pp\Pp}\xspace}
\newcommand{\pb}{\unit{pb}}
\providecommand{\cmsTable}[1]{\resizebox{\textwidth}{!}{#1}}
\newlength\cmsTabSkip\setlength{\cmsTabSkip}{2ex}

\cmsNoteHeader{TOP-22-010}
\title{Search for new Higgs bosons via same-sign top quark pair production in association with a jet in proton-proton collisions at \texorpdfstring{$\sqrt{s}=13\TeV$}{sqrt(s) = 13 TeV}}

\date{\today}

\abstract{A search is presented for new Higgs bosons in proton-proton (\pp) collision events in which a same-sign top quark pair is produced in association with a jet, via the $\pp\to \PQt\PH/\PA\to\ttc$ and $\pp\to \PQt\PH/\PA\to\ttu$ processes. Here, \PH and \PA represent the extra scalar and pseudoscalar boson, respectively, of the second Higgs doublet in the generalized two-Higgs-doublet model (g2HDM). The search is based on \pp collision data collected at a center-of-mass energy of 13\TeV with the CMS detector at the LHC, corresponding to an integrated luminosity of 138\fbinv. Final states with a same-sign lepton pair in association with jets and missing transverse momentum are considered. New Higgs bosons in the 200--1000\GeV mass range and new Yukawa couplings between 0.1 and 1.0 are targeted in the search, for scenarios in which either \PH or \PA appear alone, or in which they coexist and interfere. No significant excess above the standard model prediction is observed. Exclusion limits are derived in the context of the g2HDM.}

\hypersetup{
pdfauthor={CMS Collaboration}, 
pdftitle={Search for new Higgs bosons via same-sign top quark pair production in association with a jet in proton-proton collisions at sqrt(s) = 13 TeV},
pdfsubject={CMS},
pdfkeywords={CMS, top quarks, extra Higgs boson, Yukawa couplings}} 

\maketitle 
\section{Introduction}
Using proton-proton (\pp) collision data recorded during the CERN LHC Run~1 (2011--2012) data-taking period, the ATLAS and CMS experiments discovered a scalar boson, $\Ph_{125}$, with a mass of 125\GeV, the properties of which are so far consistent with those of the Higgs boson of the standard model (SM)~\cite{Aad:2012tfa,Chatrchyan:2012ufa,Khachatryan:2016vau}.
The analysis of the additional Run~2 (2015--2018) data enabled more precise measurements of the Higgs boson properties and further constrained the possible contributions from new physics beyond the SM~\cite{Almeida:2018cld,Almeida:2021asy}. 
The ATLAS and CMS experiments have measured the Yukawa couplings of the $\Ph_{125}$ boson to the \PQt and \PQb quarks, and the \PGt and \PGm leptons~\cite{ATLAS:2022vkf,CMS:2022dwd}. 
In all cases, the couplings are found to be consistent with the SM expectations within uncertainties. 

Having established the existence of one scalar SU(2) doublet, natural questions emerge about the possibility of additional doublets.
A second doublet is predicted in two-Higgs-doublet models (2HDM), resulting in five physical scalar bosons: {\it CP}-even neutral scalar bosons \Ph and \PH with $\mH>\mh$, a {\it CP}-odd pseudoscalar boson \PA, and two charged Higgs bosons $\PH^{\pm}$~\cite{Branco:2011iw}.
These additional Higgs bosons at mass scales ${\lesssim} 1\TeV$ may still exist within the reach of the LHC~\cite{Kohda:2017fkn}, but their signatures may be suppressed by fermion mass-mixing hierarchy~\cite{Hou:1991un} or alignment mechanisms, 
where the mixing angle $\gamma$ between the two {\it CP}-even scalar bosons \Ph and \PH has a value for which $\cos\gamma\approx0$~\cite{Hou:2017hiw}.
In the alignment limit~\cite{Gunion:2002zf,Carena:2013ooa,BhupalDev:2014bir}, the lightest scalar boson \Ph becomes the SM $\Ph_{125}$ boson, and the extra spin-0 bosons do not interact with vector boson pairs (\ie, \PH{}\PV{}\PV and \PA{}\PV{}\PV interactions do not occur, where \PV{} is a \PW or \PZ boson).
Alignment suppresses flavor changing neutral current interactions of the \Ph boson but allows them for the \PH and \PA bosons.
Therefore, the lack of a hint of the processes such as $\PQt\to\PQc\Ph_{125}/\PQu\Ph_{125}$ or $\Ph_{125}\to\PGm\PGt/\Pe\PGt$ at the LHC may be because their amplitudes are suppressed by alignment. 
As shown in Ref.~\cite{Hou:2017hiw}, the alignment scenario may occur if all extra Higgs quartic couplings are $\mathcal{O}(1)$ and no $\mathbb{Z}_2$ symmetry requirement is imposed, allowing flavor changing neutral Higgs couplings involving \PH and \PA bosons. 
This scenario without $\mathbb{Z}_2$ symmetry can be studied in the framework of the generalized 2HDM (g2HDM)~\cite{Lee:1973iz, Hou:1991un, Branco:2011iw}, with new Yukawa couplings such as $\rho_{\PGt\PGm}$, $\rho_{\PQt\PQu}$, $\rho_{\PQt\PQc}$, and $\rho_{\PQt\PQt}$ that may assume significantly large values~\cite{Lee:1973iz, Hou:1991un, Branco:2011iw}.  
These new Yukawa couplings combined with Higgs quartic couplings of $\approx{}1$ may also explain the electroweak baryogenesis~\cite{Fuyuto:2017ewj,Fuyuto:2019svr} and therefore the disappearance of antimatter soon after the Big Bang.
The new top Yukawa coupling $\rho_{\PQt\PQc}$ with a similar strength to $\rho_{\tau\mu}$ may be compatible with the observed data depending on the $\PH^\pm$ mass, which may also explain a possible muon $\text{g}-2$ anomaly~\cite{Muong-2:2023cdq}.
These arguments strongly motivate to verify with direct experimental searches whether new Yukawa couplings exist. 
  The processes discussed above have been recently explored by the ATLAS experiment~\cite{ATLAS:2023tlp}.

\begin{figure}[ht!]
  \begin{center}
        \includegraphics[width=0.4\textwidth]{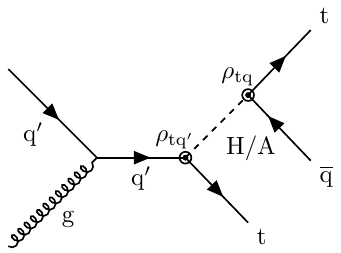}
   \caption{Representative Feynman diagram for \ttq (\PQq = \PQu, \PQc) production through a new scalar (\PH) or pseudoscalar (\PA) Higgs boson. 
   In this analysis, events with $\PQq=\PQq'$ are considered.
}
    \label{fig:feyn_ttc}
  \end{center}
\end{figure}

In this Letter, we present a search for the existence of the real part of these couplings, $\rho_{\PQt\PQu}$ and $\rho_{\PQt\PQc}$, 
through the $\pp\to\PQt\PH/\PA\to\ttq~(\PQq = \PQu,\PQc)$ and the charge conjugate processes, 
considering one coupling at a time and keeping the other at zero, using the Run 2 data set recorded by the CMS experiment.  
 A representative Feynman diagram for the \ttq process is displayed in Fig.~\ref{fig:feyn_ttc}. 
The final-state signature for \ttq explored in this Letter consists of two same-sign (SS) leptons with at least three jets, two of which are identified as \PQb jets and one compatible with originating from a \PQu or \PQc quark, and missing transverse momentum. 
The main background consists of events with nonprompt leptons originating from leptonic decays of heavy quarks, hadrons misidentified as leptons, electrons from photon conversions, or jets misidentified as leptons. 

The analysis is performed using \pp collision data at a center-of-mass energy of 13\TeV.
The data were collected with the CMS detector at the LHC between 2016 and 2018, and correspond to an integrated luminosity of 138\fbinv.
Tabulated results are accessible from the HEPData record for this analysis~\cite{hepdata}.

The Letter is organized as follows: Section~\ref{sec:CMSdet} describes the CMS detector, along with object reconstruction. 
In Section~\ref{sec:sim}, a description of the simulation samples used in the analysis is given. 
The event selection and reconstruction procedure of the physics objects are given in Section~\ref{sec:eventsel}.
Following, in Section~\ref{sec:signalextract}, the signal extraction procedure is explained. 
Section~\ref{sec:systematics} discusses the sources of systematic uncertainties.
The results of the analysis are presented in Section~\ref{sec:results}. 
Finally, a summary of the analysis is given in Section~\ref{sec:summary}.

\section{The CMS detector and object reconstruction}
\label{sec:CMSdet}
 The CMS apparatus~\cite{CMS:2008xjf} is a multipurpose, nearly hermetic detector, designed to trigger on~\cite{CMS:2020cmk,CMS:2016ngn} and identify electrons, muons, photons, and (charged and neutral) hadrons~\cite{CMS:2020uim,Sirunyan:2018,CMS:2014pgm}. 
 A global ``particle-flow" (PF) algorithm~\cite{CMS:2017yfk} aims to reconstruct all individual particles in an event, combining information provided by the all-silicon inner tracker and by the crystal electromagnetic (ECAL) and brass-scintillator hadron calorimeters (HCAL), operating inside a 3.8\unit{T} superconducting solenoid, with data from the gas-ionization muon detectors embedded in the flux-return yoke outside the solenoid. 
 
 The primary vertex is taken to be the vertex corresponding to the hardest scattering in the event, evaluated using tracking information alone, as described in Section~9.4.1 of Ref.~\cite{CMS-TDR-15-02}. 
The energy of photons is obtained from the ECAL measurement. The energy of electrons is determined from a combination of the electron momentum at the primary interaction vertex as determined by the tracker, the energy of the corresponding ECAL cluster, and the energy sum of all bremsstrahlung photons spatially compatible with originating from the electron track. The energy of muons is obtained from the curvature of the corresponding track. 
The energy of charged hadrons is determined from a combination of their momentum measured in the tracker and the matching ECAL and HCAL energy deposits, corrected for the response function of the calorimeters to hadronic showers. Finally, the energy of neutral hadrons is obtained from the corresponding corrected ECAL and HCAL energies. 

Jets are built from PF particles using the anti-\kt algorithm~\cite{Cacciari:2008gp,Cacciari:2011ma} with a distance parameter of 0.4. Jet momentum is determined as the vectorial sum of all particle momenta in the jet, and is found from simulation to be, on average, within 5 to 10\% of the true momentum over the whole \pt spectrum and detector acceptance. Additional \pp interactions within the same or nearby bunch crossings can contribute additional tracks and calorimetric energy depositions, increasing the apparent jet momentum. 
To mitigate this effect, tracks identified to be originating from the multiple pp interactions occurring in the same or nearby bunch crossings (pileup) are discarded and an offset correction is applied to correct for remaining contributions~\cite{CMS:2020ebo}. Jet energy corrections are derived from simulation studies so that the average measured energy of jets becomes identical to that of particle level jets. In situ measurements of the momentum balance in dijet, $\text{photon+jet}$, \zjet, and multijet events are used to determine any residual differences between the jet energy scale in data and in simulation, and appropriate corrections are made~\cite{CMS:2016lmd}. Additional selection criteria are applied to each jet to remove jets potentially dominated by instrumental effects or reconstruction failures~\cite{CMS:2020ebo}. 
The missing transverse momentum vector \ptvecmiss is computed as the negative vector sum of the transverse momenta of all the PF candidates in an event, and its magnitude is denoted as \ptmiss~\cite{CMS:2019ctu}. The \ptvecmiss is modified to account for corrections to the energy scale of the reconstructed jets in the event. 

Events of interest are selected using a two-tiered trigger system. The first level (L1), composed of custom hardware processors, uses information from the calorimeters and muon detectors to select events at a rate of around 100\unit{kHz} within a fixed latency of about 4\mus~\cite{CMS:2020cmk}. The second level, known as the high-level trigger, consists of a farm of processors running a version of the full event reconstruction software optimized for fast processing, and reduces the event rate to around 1\unit{kHz} before data storage~\cite{CMS:2016ngn}. 

\section{Simulated event samples}
\label{sec:sim}
The signal and background processes are simulated using several Monte Carlo (MC) programs.
The \ttq signal samples are simulated at leading order (LO) in perturbative quantum chromodynamics (QCD) with up to 
two additional noncollinear high-\pt partons
using \MGvATNLO 2.6.5 ~\cite{Alwall:2014hca}, and the hard-process simulation is interfaced with parton shower jets using the MLM~\cite{Alwall:2007fs} matching prescription.
Different signal samples are simulated for the case where a mass between 200 and 1000\GeV is assumed for \PA (\PH), and a mass too high to be of reach for our search is assumed for \PH (\PA). 
For these signal samples, there is no interference between \PA and \PH. 
Signal samples are also generated for the case where both \PA and \PH coexist and interfere assuming a mass difference $\mA-\mH = 50\GeV$, following Ref.~\cite{Hou:2020chc}.
The amplitudes for the processes $\PQq\Pg\to\PQt\PA\to\ttq$ and $\PQq\Pg\to\PQt\PH\to\ttq$ cancel when \PA and \PH are mass degenerate. The signal cross section is not significantly modified with respect to the noninterference case when the mass differences are larger than ${\approx}100\GeV$. 
Therefore, the investigated mass difference is chosen to retain some interference effects along with reasonably high cross sections. 
Depending on the coupling value and the interference assumption we consider in this Letter, the cross sections for the \ttu process range from ${\approx}1\times10^{-4}$ to ${\approx}7\times10^{-1}\pb$  and for \ttc range from ${\approx}5\times10^{-6}$ to ${\approx}7\times10^{-2}\pb$, as calculated with \MGvATNLO{} at LO precision in QCD. 
The inclusive cross section for the \ttq process does not change, within uncertainties, when the intermediary particle is \PH or \PA, when there is no \PA-\PH interference.   

Background events generated by the \MGvATNLO generator are \ttV, \vjet, \ttVV, \ttVH, 
\tZ, \tttj, \tttW, \tttt, \wz, vector boson scattering (VBS, \ie, \WLLJJ or $\PV^{\pm}\PV^{\pm}$JJ), \VVV, and single top quark $s$-channel production. 
Among these processes, FxFx matching~\cite{Frederix:2012ps} is used for \ttW, \zjet, and the QCD part of \wz production, while MLM matching~\cite{Alwall:2007fs} is used for \wjet. 
The \POWHEG~2.0 generator~\cite{Nason:2004rx,Frixione:2007vw,Alioli:2010xd} is used to simulate events at next-to-LO  in perturbative QCD for \ttbar~\cite{Frixione:2007nw},
\zz and $\PW^\pm\PW^\mp$~\cite{Melia:2011tj}, single top quark \tW~\cite{Re:2010bp} and $t$-channel~\cite{Alioli:2009je}, and \ttH production~\cite{Hartanto:2015uka}.  

Both the signal and background samples are interfaced with \PYTHIA v8.240~\cite{Sjostrand:2014zea} for parton showering, fragmentation, and hadronization.
The underlying event is also modeled with \PYTHIA, using the CP5 tune~\cite{Sirunyan:2019dfx}.
The parton distribution functions (PDFs) are taken from the NNPDF3.1 next-to-next-to-LO set~\cite{NNPDF:2017mvq}.
Finally, for both signal and background MC event samples, the detector response is modeled with the \GEANTfour package \cite{Agostinelli:2002hh}.
The simulated events are weighted to match the observables sensitive to pileup to those observed in the data~\cite{CMS:2020ebo}.

The simulated events are corrected for the differences between data and simulation of the trigger efficiencies, lepton identification, and lepton isolation. 
The corrections are determined using the ``tag-and-probe'' method applied on Z boson candidate events~\cite{CMS:2011aa} and are parametrized as functions of the lepton \pt and pseudorapidity ($\eta$). 

\section{Event reconstruction and selection}
\label{sec:eventsel}
Data are selected online with a combination of double lepton triggers and single lepton triggers to enhance signal efficiency.
Charged leptons (electrons or muons) are required to originate from the primary vertex to reject contributions from pileup.
Identification of electrons makes use of a multivariate discriminant~\cite{CMS:2020mpn,CMS:2022tkv} combining observables sensitive to the
matching of charged-particle tracks in the tracker to the energy deposits in the ECAL and the
amount of bremsstrahlung photons emitted along their trajectory~\cite{CMS:2020uim}. 
The number of electrons originating from asymmetric photon conversions is reduced by requiring that 
their associated track has no missing hits in the innermost layers of the silicon tracker. 
To ensure the correct assignment of the electric charge of the electrons, further selection criteria were applied~\cite{Khachatryan:2015hwa}. 
The separation of prompt leptons from nonprompt leptons reconstructed within jets is achieved by using an isolation variable, relative to the lepton \pt. The isolation variable is constructed from the scalar \pt sum of charged hadrons, neutral hadrons, and photons reconstructed within a cone of variable size that depends on the lepton \pt~\cite{Rehermann:2010vq}.
``Loose'' leptons are those that pass the kinematic selection criteria of $\pt>10\GeV$ and $\abs{\eta}<2.5~(2.4)$ for electrons (muons).   
Electrons that fall in the region between the ECAL barrel and endcaps ($1.442<\abs{\eta}<1.556$) are vetoed. 
``Tight'' leptons are in addition required to have $\pt>20\GeV$. 
A multivariate discriminant is furthermore used to separate prompt leptons and nonprompt leptons 
originating from the decays of hadrons inside jets, hadrons misidentified as leptons, and electrons from photon conversions. 
The jets must have $\pt>30\GeV$ and $\abs{\eta}<2.4$ to be considered. 
In addition, they are required to be separated from identified leptons by $\Delta R = \sqrt{\smash[b]{({\Delta}\eta)^2+({\Delta}\phi)^2}}>0.4$, 
where $\phi$ is the azimuthal angle measured in radians.
At least three jets, and $\ptmiss>30\GeV$ are required in each event. 
Jets from the hadronization of \PQb quarks (\PQb jets), \PQc quarks (\PQc jets), and light (\ie, light-quark or gluon) jets are distinguished through the ratios of the corresponding jet flavor probabilities, defined as CvsL (\PQc vs light quark) and CvsB (\PQc vs \PQb quark)~\cite{CMS:2021scf,CMS-DP-2023-006}, respectively. 
These probabilities are obtained  from the \textsc{DeepJet}~\cite{Bols:2020bkb,CMS-DP-2023-005} algorithm: a neural network utilizing global variables, charged, neutral particle and secondary vertex features in a jet to perform flavor tagging.
The differences of the shapes of the CvsL and CvsB discriminant distributions between data and simulation are corrected by flavor-dependent scale factors. 

\begin{figure}[ht!]
  \centering
   \includegraphics[width=0.49\textwidth]{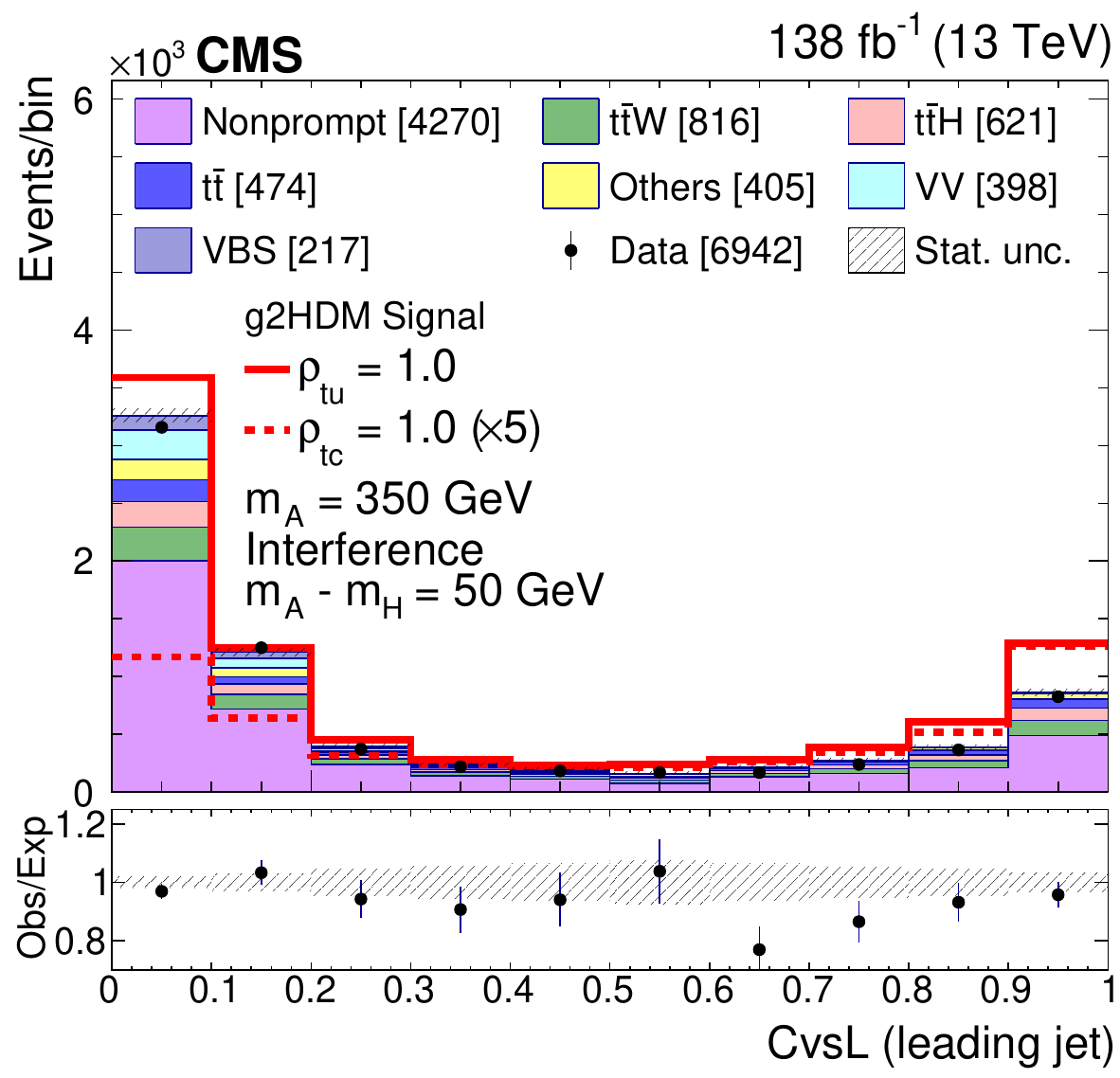}
   \includegraphics[width=0.49\textwidth]{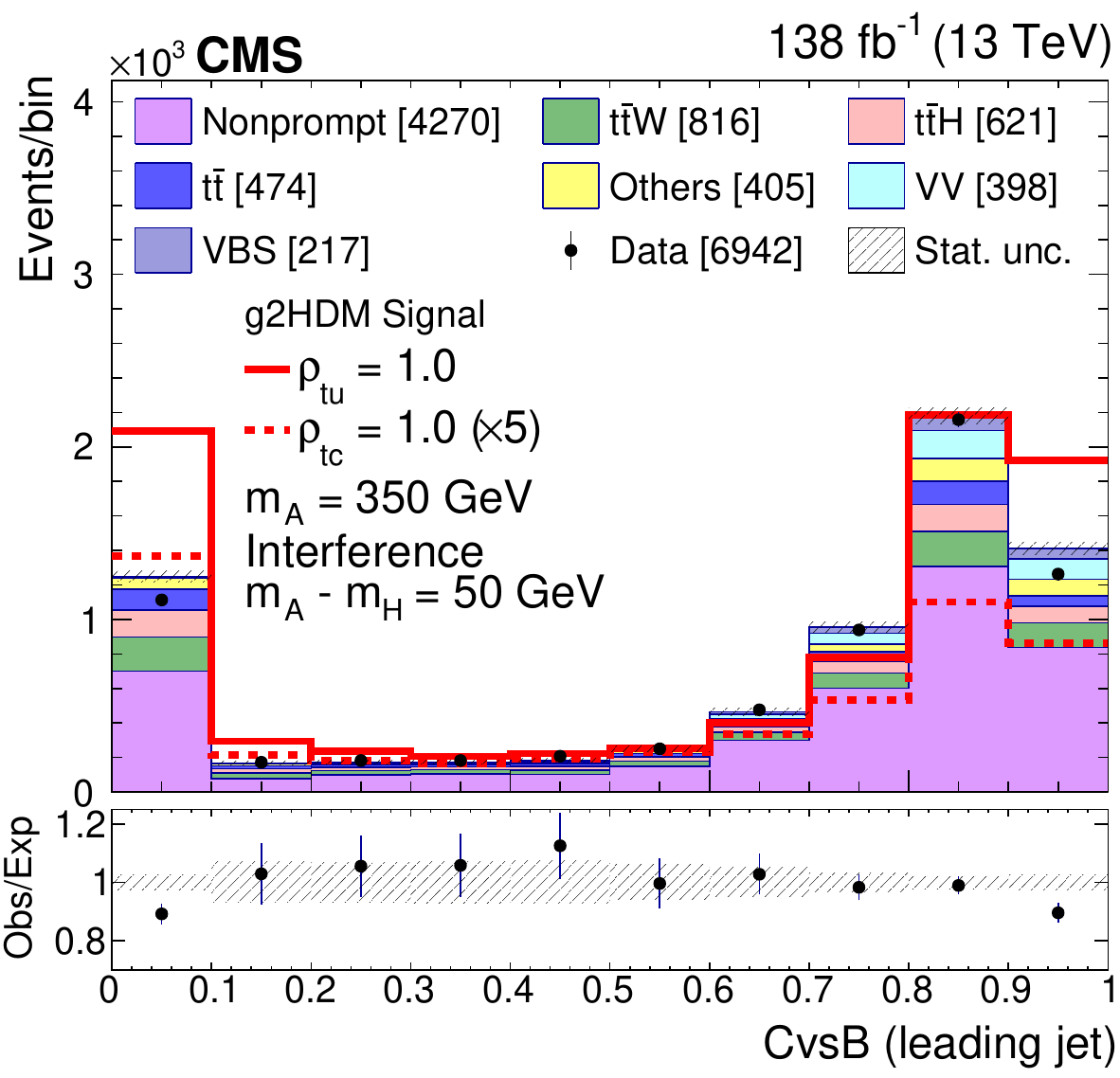}
  \caption{The pre-fit CvsL (\cmsLeft) and CvsB (\cmsRight) distributions for the selected highest \pt jet. The predictions for $\mA = 350\GeV$ with \PA-\PH interference assuming $\mA -\mH=50\GeV$ for $\rho_{\PQt\PQu}=1.0$ (solid red line) and $\rho_{\PQt\PQc}=1.0$ (dashed red line) are also displayed. The numbers in square brackets represent the yields for each sample. The uncertainty bars on the points and the hatched bands represent the statistical uncertainties in the data and in the background predictions, respectively. 
  Beneath each plot the ratio of data to predictions is shown. The uncertainty bars in the ratio plots include statistical uncertainties in the data and in the background predictions.}
\label{fig:cvslb}
\end{figure}

The search is performed in three SS dilepton categories: \ee, \mumu, and \emu.
The event selection ensures complete orthogonality of these three categories to avoid any possible double counting of events selected in the analysis.
Events are required to have two tight leptons with the same electric charge for which the leading lepton has $\pt>30\GeV$ and subleading lepton $\pt>20\GeV$.
Events containing a third lepton are vetoed if the additional lepton  passes the loose selection.
The two leptons are required to be separated by $\Delta R(\ell_1,\ell_2)>0.3$ and required to have an invariant mass $m(\ell_1,\ell_2)>20\GeV$. 
Events with $60<m(\ell_1,\ell_2)<120\GeV$ in the \ee channel are vetoed to suppress the Drell-Yan (DY) background contribution, the main source of charge misidentified leptons. 
This veto reduces the selected number of DY events in simulation by about 94\% after the aforementioned requirements. 
This veto is not applied to the \emu and \mumu channels, as the charge misidentification rate for muons is negligible. 
The CvsL and CvsB distributions for the leading jet obtained from simulated events and data are displayed in Fig.~\ref{fig:cvslb}. 

The nonprompt-lepton background is estimated from control regions in the data using the fake-factor method~\cite{CMS:2018fdh}. 
A dijet enriched sample is selected from data, from which the probability $f_p$ for a loose nonprompt lepton (originating from a jet) to also pass the tight lepton selection criteria is estimated in bins of \pt and $\abs{\eta}$.
The nonprompt-lepton background contribution is then estimated by applying 
$f_p$ to events in a data control sample in which at least one lepton fails the tight lepton selection criteria.
A data-driven approach is followed to evaluate the lepton charge misidentification rate in the \ee channel using DY events with same- and opposite-sign electric charges.
It is estimated in bins of lepton \pt and  $\eta$, and is used to calibrate simulated events. 
The rate, which increases with \pt and $\eta$, varies from ${\approx}10^{-5}$ to ${\approx}10^{-2}$.  

\section{Signal extraction}
\label{sec:signalextract}
A boosted decision tree (BDT) discriminant~\cite{Friedman:2001wbq} is used to separate signal and background events. 
We train the BDT model in the signal region, where process yields of both backgrounds and signal are small.
For the training, we combine the inputs from different channels.
By comparing the BDT output distributions for the training and test samples, we found no evidence of overtraining. 
We use an optimized set of input observables, listed in Table~\ref{tab:bdt_inputs}, for our signal topology.
Flavor tagging variables are used to differentiate \PQb quarks originating from the top quark decay, 
and to differentiate \PQc quark and light-quark jets. 
Pairwise kinematic variables related to jets are incorporated that are sensitive to the resonance structures, 
specifically to top quarks and \PA/\PH. 
Other kinematic features, such as \HT and \ptmiss, are also sensitive to our signal and are included. 
To mitigate the impact of resonances such as the DY process, the dilepton (+jets) invariant mass is also used as an input to the BDT.

\begin{table*}[!htpb] 
\centering 
\topcaption{Input variables of the BDT. Jets and leptons are ordered by \pt. } 
\begin{tabular}{lll} 
  \multicolumn{2}{c}{Input variables of the BDT} &  \\
\hline
CvsL$(j_a)$ & $a=1,2,3$ & Charm- vs light-quark jet identification variable\\ 
CvsB$(j_a)$ & $a=1,2,3$ & Charm- vs bottom-quark jet identification variable\\
$\Delta R(j_a,j_b)$ & $1\leq a<b\leq3$ & Angular separation between jets\\  
$m(j_a,j_b)$ & $1\leq a<b\leq3$ & Invariant mass of jet pairs\\
$\Delta R(j_a,l_b)$ & $a=1,2,3$; $b=1,2$ & Angular separation between jet and lepton\\ 
$m(j_a,l_b)$ & $a=1,2,3$; $b=1,2$ & Invariant mass of jet-lepton pairs\\
$\pt(\ell_a)$ & $a=1,2$ & Transverse momentum of leptons\\ 
$m(\ell_1,\ell_2,j_a)$ & $a=1,2,3$ & Invariant mass of the two leptons plus the highest \pt jet\\
$m(\ell_1,\ell_2)$ & & Invariant mass of the two leptons\\ 
\HT & & Scalar \pt sum of the jets\\ 
\ptmiss & & Missing transverse momentum\\
\end{tabular} 
\label{tab:bdt_inputs} 
\end{table*}

We use half of the available simulated events for the BDT training. 
The main backgrounds in the signal region are \ttbar (fully leptonic decay of the top quark-antiquark pair where one lepton has a misidentified charge), nonprompt leptons (dominated by semileptonic \ttbar decay), \ttW, and \ttH.
In the signal extraction, we use events from data to estimate the nonprompt-lepton background contribution, but we do not use these events for training. 
Instead, we use simulated samples of semileptonically decaying \ttbar events for the BDT training, to mimic the misidentified lepton in the nonprompt-lepton background.
We do not include events from \wjet  or  single top quark simulation, due to a limited number of simulated events. 
The various backgrounds are added corresponding to their cross sections.
The MC event samples used for the BDT training are 
\ttc or \ttu signal,
semileptonic \ttbar,
dileptonic \ttbar (\ttbar),
\ttV,
\VV,
\VVV,
\ttH,
\ttVH,
\ttVV,
\tttj, \tttW, \tttt, and
VBS.
Three values for the couplings $\rho_{\PQt\PQu}$ and $\rho_{\PQt\PQc}$, namely, 0.1, 0.4, and 1.0 are considered.
For each coupling, ten \mA points are considered: 200, 300, 350, 400, 500, 600, 700, 800, 900, and 1000\GeV.
Using the same coupling assumptions, we also consider the more realistic case with the \PA-\PH interference assuming a fixed mass difference $\mA -\mH=50\GeV$, with the following nine \mA-\mH combinations; 250-200, 300-250, 350-300, 400-350, 550-500, 700-650, 800-750, 900-850, and 1000-950\GeV.
The signal extraction is performed using the shape of the BDT discriminant distributions. 
Because of the kinematic differences and different background compositions, the BDTs are trained independently for four data-taking periods, two in 2016, and one for each of the years 2017 and 2018; for each mass assumption, namely for each \mA either with no \PH present or with \PA-\PH interference; and for $\rho_{\PQt\PQu}=0.4$ and $\rho_{\PQt\PQc}=0.4$.
Therefore, a total of 152 BDTs are trained.
The values of the BDT hyperparameters, as defined in the \textsc{TMVA} package~\cite{Hocker:2007ht}, are chosen after scanning a range of
values, avoiding overtraining, and requiring a high background rejection rate while maintaining a high signal efficiency.  
We use these same model parameters for all three years of data taking.
The BDTs offer high background rejection, while simultaneously maintaining a high signal efficiency. 
For instance, when considering $\rho_{\PQt\PQc} = 0.4$, and a signal efficiency of 96\%, the background rejection rates for $\mA = 200\GeV$ and $1000\GeV$ are 50\% and 76\%, respectively. 

{\tolerance=400
The signal strength parameter, $\mu$, is obtained using a simultaneous maximum likelihood fit~\cite{CMS-NOTE-2011-005}, in analyses performed for three decay modes and for four different eras (\ie 12 categories) for each signal mass-coupling assumption independently.
It is defined as $\mu=\sigma/\sigma_{\mathrm{theory}}$, where $\sigma_{\mathrm{theory}}$ is
the predicted \ttq production cross section in g2HDM calculated with \MGvATNLO+\PYTHIA, and  $\sigma$ is the upper limit on the observed cross section.
To ensure numerical stability, the bin width is chosen to have a high enough yield in each bin of the BDT distributions.
The BDT distribution ranges from $-1$ to 1, but to reduce the impact on the fit of background-dominated regions, events are selected with a BDT score ${>}-0.6$. This requirement on the BDT score also helps to improve the stability of the fit and the corresponding fit uncertainties. 
Except for the dominant nonprompt-lepton background, the predicted yields and the shapes of the BDT distributions are taken directly from the simulation.
When performing the maximum likelihood fit, individual templates are used for the VBS, \VV, \ttbar (fully leptonic decays), \ttH, \ttW, and nonprompt-lepton background processes, while a combined template labeled {\it others} uses a combination of background processes, namely: DY, \ttVV, \ttVH, \ttZ, \tttj, \tttW, \tttt, \VVV, \tZq, \tW, and \tbarW. 
\par}

\section{Systematic uncertainties}
\label{sec:systematics}
Systematic uncertainties, arising from various sources, such as detector effects, theoretical uncertainties, and mismodeling,
are evaluated and categorized in two main groups: experimental uncertainties and modeling uncertainties.   
Several effects modify the event yields or shapes of the measured distributions. 
The systematic uncertainties are modeled as nuisance parameters in the maximum likelihood estimation procedure used for determining the best fit signal strength $\hat{\mu}$. 
The nuisance parameters are allowed to modify the shape and normalization of the backgrounds and signal in the fit using the method in Ref.~\cite{CMS-NOTE-2011-005}. 
The chosen probability density function (pdf) used for the interpolation between nominal and systematic templates is dependent on whether the associated nuisance parameter can also affect the shape of the nominal template or not. Log-normal pdfs are used for nuisance parameters that only affect the distributions normalisation, whereas Gaussian distributions are used for those that can affect the shape as well.
Experimental uncertainties, where appropriate, and all modeling uncertainties are taken as correlated across years and channels. 
The effect of the limited number of simulated events is estimated using the Barlow--Beeston-lite method~\cite{Barlow:1993dm}. 

The experimental uncertainties are those related to the integrated luminosity~\cite{CMS:2021xjt,CMS-PAS-LUM-17-004,CMS-PAS-LUM-18-002},
pileup, L1 trigger inefficiency, nonprompt-lepton background estimation, jet energy scale and resolution, 
unclustered energy scale, lepton identification and trigger efficiencies, charge misidentification, muon momentum scale, and heavy-quark, light-quark and gluon jet identification.
The dominant experimental uncertainty sources are related to the flavor tagger discriminant shape calibration, and nonprompt-lepton background estimation. 
The modeling uncertainty sources are matrix-element renormalization and factorization scales, parton shower scales that control the initial- and final-state radiation, and PDFs. For the signal, these are included as separate nuisance parameters that affect the shape. 
For the backgrounds estimated using MC event samples, nuisance parameters are assumed to only affect the yields, and the corresponding uncertainties in the theoretical cross section calculations are used. 
The most significant modeling uncertainty in the analysis is attributed to the normalization of the \ttW background, regardless of the \PA-\PH interference. 
The contributions of all uncertainty sources on the pre-fit expected event yields are displayed in Table~\ref{tab:systematics}, for $\rho_{\PQt\PQc}=0.4$ and  $\mA = 350\GeV$ with \PA-\PH interference assuming $\mA -\mH=50\GeV$.
The uncertainties in other scenarios are similar.

\begin{table*}[!htpb]
\centering
\topcaption{Summary of systematic uncertainties for $\rho_{\PQt\PQc}=0.4$ and $\mA = 350\GeV$ with \PA-\PH interference assuming $\mA -\mH=50\GeV$. The first column indicates the source of uncertainty. The second column specifies whether the shape of the fit discriminant is affected by the nuisance parameter (\checkmark) or not (dash). 
The impact in percent of these nuisance parameters on the pre-fit expected event yields is displayed in the third column. This column is subdivided into three event categories representing the analysis channels. The percentage impacts are given as a range of values representing the minimum and maximum differences obtained in the different bins of the BDT distribution through the four data-taking periods. The numbers for the normalization component of the nonprompt lepton background represent the uncertainties used for each data-taking period. Whether or not a nuisance parameter is taken correlated across years and categories is specified in the last two columns. The luminosity and jet flavor identification nuisances are only partially correlated across years.}
\begin{tabular}{lccccccc}
  & & \multicolumn{3}{c}{Category} & \multicolumn{2}{c}{Correlated across}\\
  Uncertainty source & Shape & \ee & \mumu & \emu & Years & Categories \\
  \hline
  \multicolumn{7}{c}{Experimental}\\
  Luminosity & \NA & 1.2--2.5 & 1.2--2.5 & 1.2--2.5 & \checkmark & \checkmark \\
  Pileup & \checkmark & $<$0.1--2.8 & $<$0.1--1.8 & $<$0.1--2.3 & \checkmark & \checkmark \\
  Trigger efficiency & \checkmark & 0.4--2.6 & 0.2--1.1 & 0.3--1.2 & \NA & \NA \\
  L1 trigger inefficiency & \checkmark & 0.1--0.8 & 0.1--0.3 & 0.1--0.4 & \checkmark & \checkmark \\
  Lepton identification & \checkmark & 0.1--1.7 & $<$0.1--0.4 & $<$0.1--0.6 & \NA & \checkmark \\
  Lepton energy scale & \checkmark & \NA & $<$0.1--0.2 & $<$0.1--0.2 & \NA & \checkmark \\
  Charge misid. & \checkmark & 1.2--13.1 & \NA & \NA & \NA & \NA \\
  Jet energy scale & \checkmark & $<$0.1--4.5 & $<$0.1--1.7 & $<$0.1--1.5 & \checkmark & \checkmark \\
  Jet energy resolution & \checkmark & $<$0.1--2.6 & $<$0.1--1.8 & $<$0.1--1.6 & \NA & \checkmark \\
  Unclustered energy  & \checkmark & $<$0.1--2.6 & $<$0.1--0.5 & $<$0.1--0.8 & \NA & \checkmark \\
  Jet flavor identification & \checkmark & $<$0.1--12.1 & $<$0.1--8.8 & $<$0.1--11.6 & \checkmark & \checkmark \\
  Nonprompt lepton BG  & & & & & & \\
  statistical component & \checkmark & $<$0.1--27.2 & 1.9--16.2 & 3.0--13.2 & \NA & \checkmark \\
  Nonprompt lepton BG & \NA & 27,15,11,10 & 27,15,11,10 & 27,15,11,10 & \NA & \checkmark \\
  \multicolumn{7}{c}{Theoretical}\\
  Signal QCD scales & \checkmark & 10.3--10.5 & 10.0--10.2 & 9.9--10.0 & \checkmark & \checkmark \\
  Signal PDF & \checkmark & 0.7 & 0.6--0.7 & 0.5--0.6 & \checkmark & \checkmark \\
  Signal parton shower & \checkmark & 3.6--4.3 & 4.0--4.3 & 6.3--7.3 & \checkmark & \checkmark \\
  \ttbar & \NA & 6.1 &  6.1 & 6.1 & \checkmark & \checkmark \\
  \VV & \NA & 4.5 & 4.5 & 4.5& \checkmark & \checkmark \\
  VBS & \NA & 10.4 & 10.4 & 10.4 & \checkmark & \checkmark \\
  \ttH & \NA & 7.8 & 7.8 & 7.8 & \checkmark & \checkmark \\
  \ttW & \NA & 10.7 & 10.7 & 10.7 & \checkmark & \checkmark \\
  Other backgrounds & \NA & 5.4 & 5.4 & 5.4 & \checkmark & \checkmark \\
\end{tabular}
\label{tab:systematics}
\end{table*}

\section{Results}
\label{sec:results}
\begin{figure}[htb!]
\centering
  \includegraphics[width=0.49\textwidth]{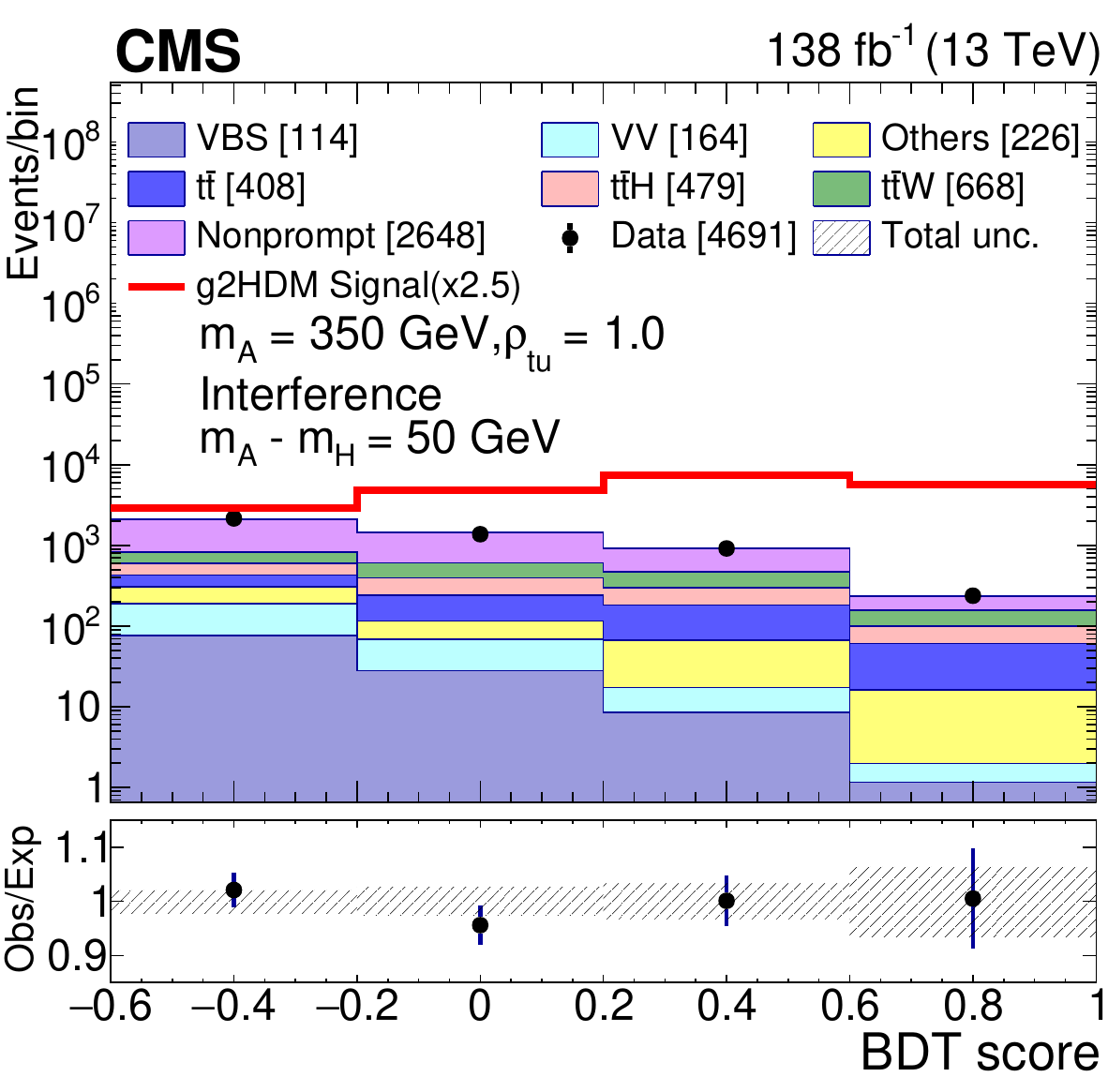}
  \includegraphics[width=0.49\textwidth]{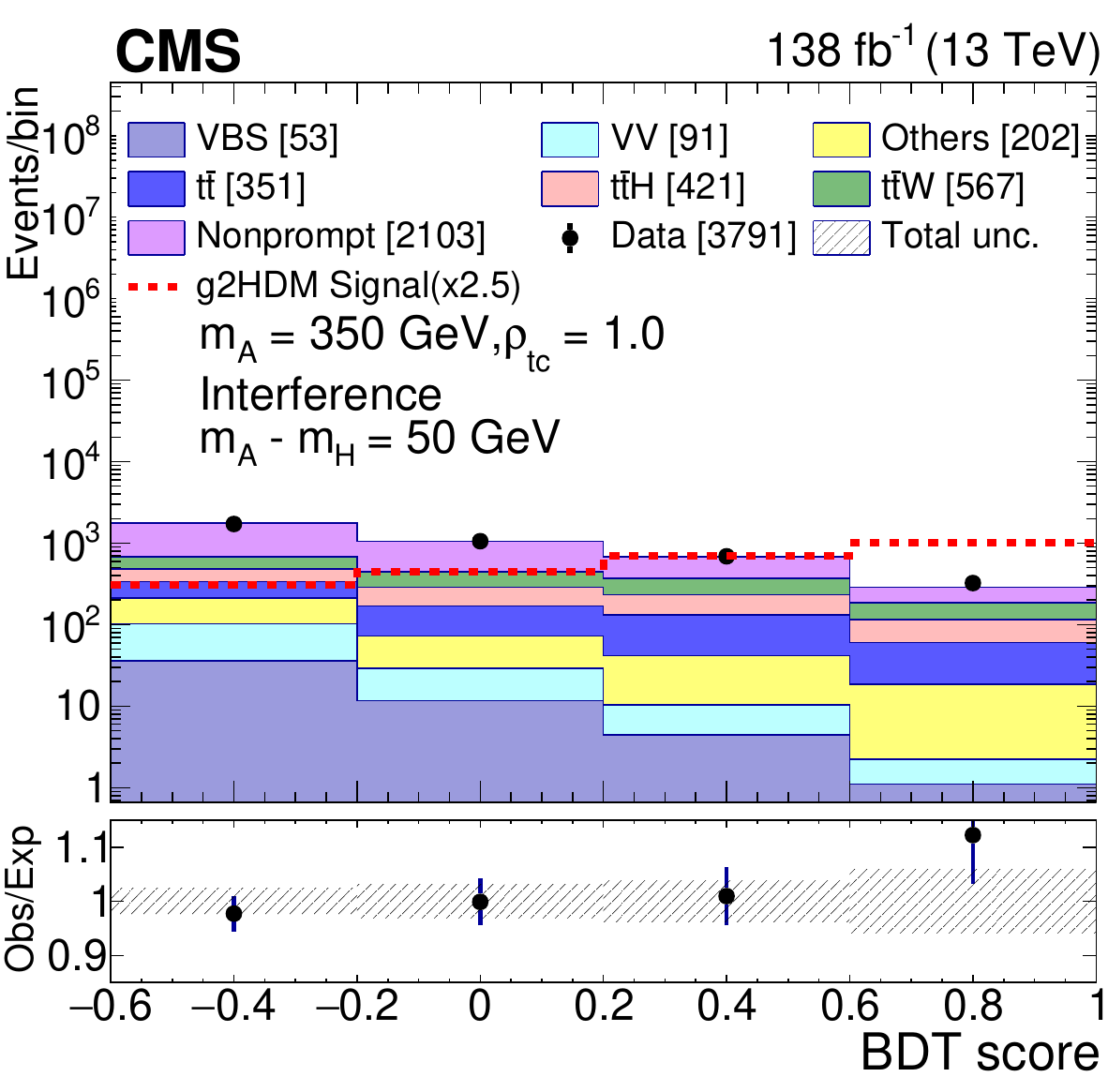}
  \caption{Post-fit distributions of the BDT discriminants combining the categories \ee, \mumu, and \emu, and all data-taking periods, for $\mA=350\GeV$ with $\rho_{\PQt\PQu}=1.0$ (\cmsLeft), and $\rho_{\PQt\PQc}=1.0$ (\cmsRight) with the \PA-\PH interference. The numbers in square brackets represent the yields for each sample. 
The uncertainty bars on the points represent the statistical uncertainties in the data. 
Beneath each plot the ratio of data to predictions is shown.
The uncertainty bars in the ratio plots include statistical uncertainties in the data and the total uncertainty in the background predictions, and the hatched bands represent the total uncertainty in the background predictions. 
}
  \label{fig:postfit_run2_rtc0p4}
\end{figure}

Figure~\ref{fig:postfit_run2_rtc0p4} displays the post-fit BDT discriminant distributions used for the signal extraction combining the categories \ee, \mumu, and \emu, and for $\rho_{\PQt\PQu}=1.0$ and $\rho_{\PQt\PQc}=1.0$ with \PA-\PH interference, assuming $\mA=350\GeV$.

\begin{figure}[!ht]
  \centering
  \includegraphics[width=0.49\textwidth]{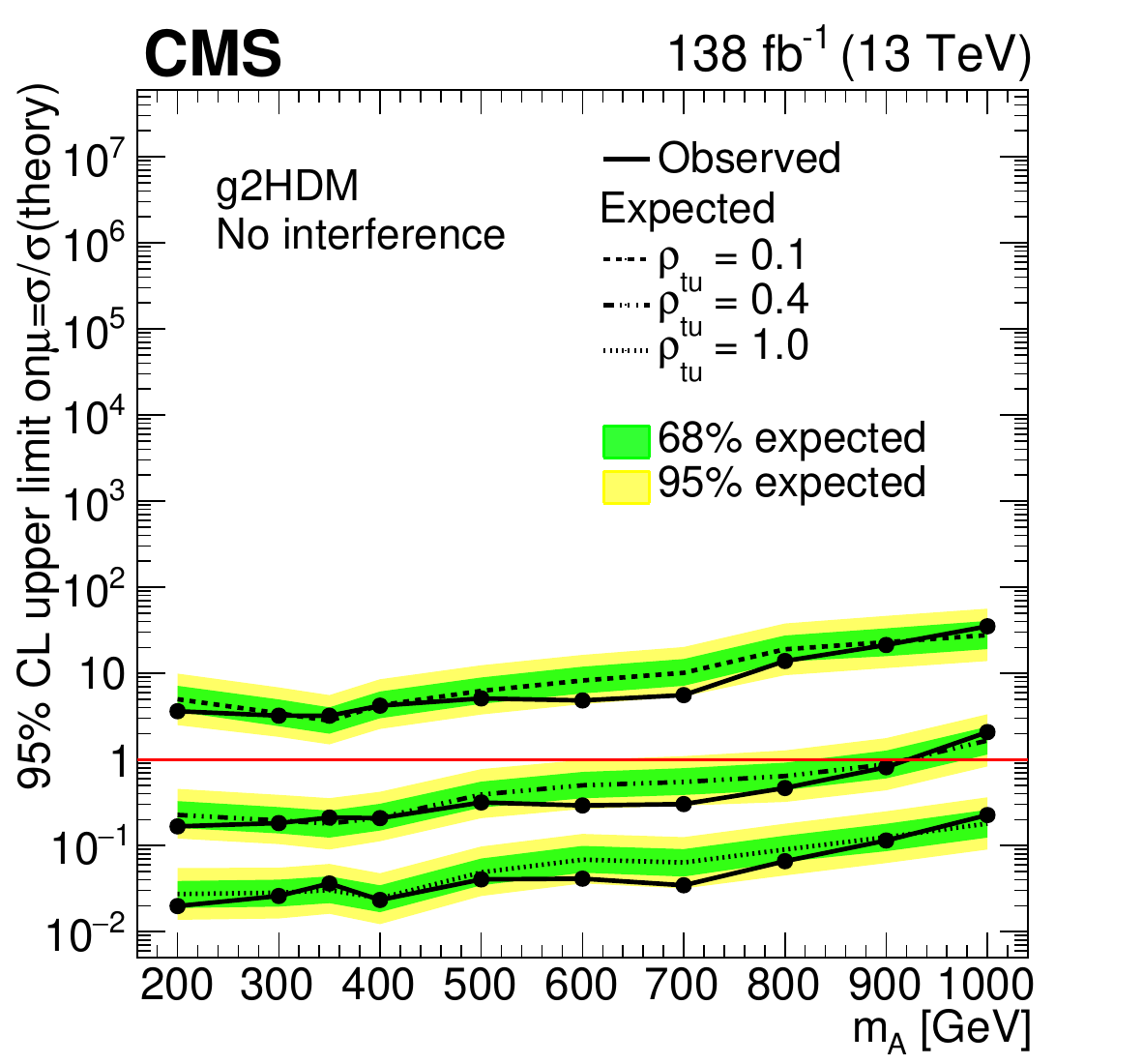}
  \includegraphics[width=0.49\textwidth]{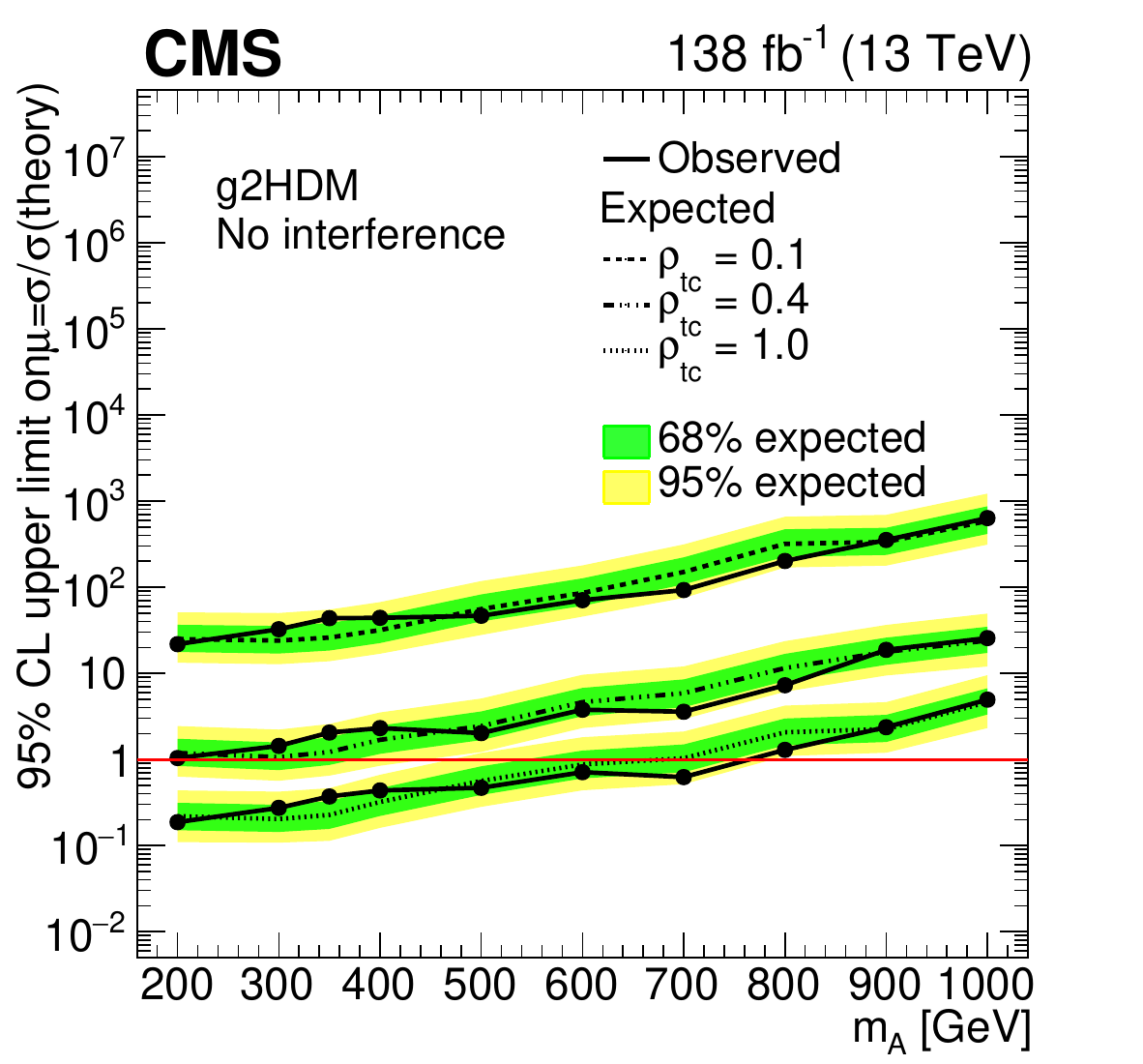}
  \caption{Observed and expected 95\% \CL upper limits on the signal strength as functions of \mA for g2HDM using different coupling assumptions: $\rho_{\PQt\PQu}$ = 0.1, 0.4, 1.0 (\cmsLeft) and $\rho_{\PQt\PQc}$ = 0.1, 0.4, 1.0 (\cmsRight) without interference, for the combination of the \ee, \mumu, and \emu categories and all data-taking periods. The inner (green) band and the outer (yellow) band indicate the regions containing 68 and 95\%, respectively, of the distribution of limits expected under the background-only hypothesis.}
  \label{fig:limit_run2_rtu_rtc_wo_interference}
\end{figure}

\begin{figure}[!ht]
  \centering
  \includegraphics[width=0.49\textwidth]{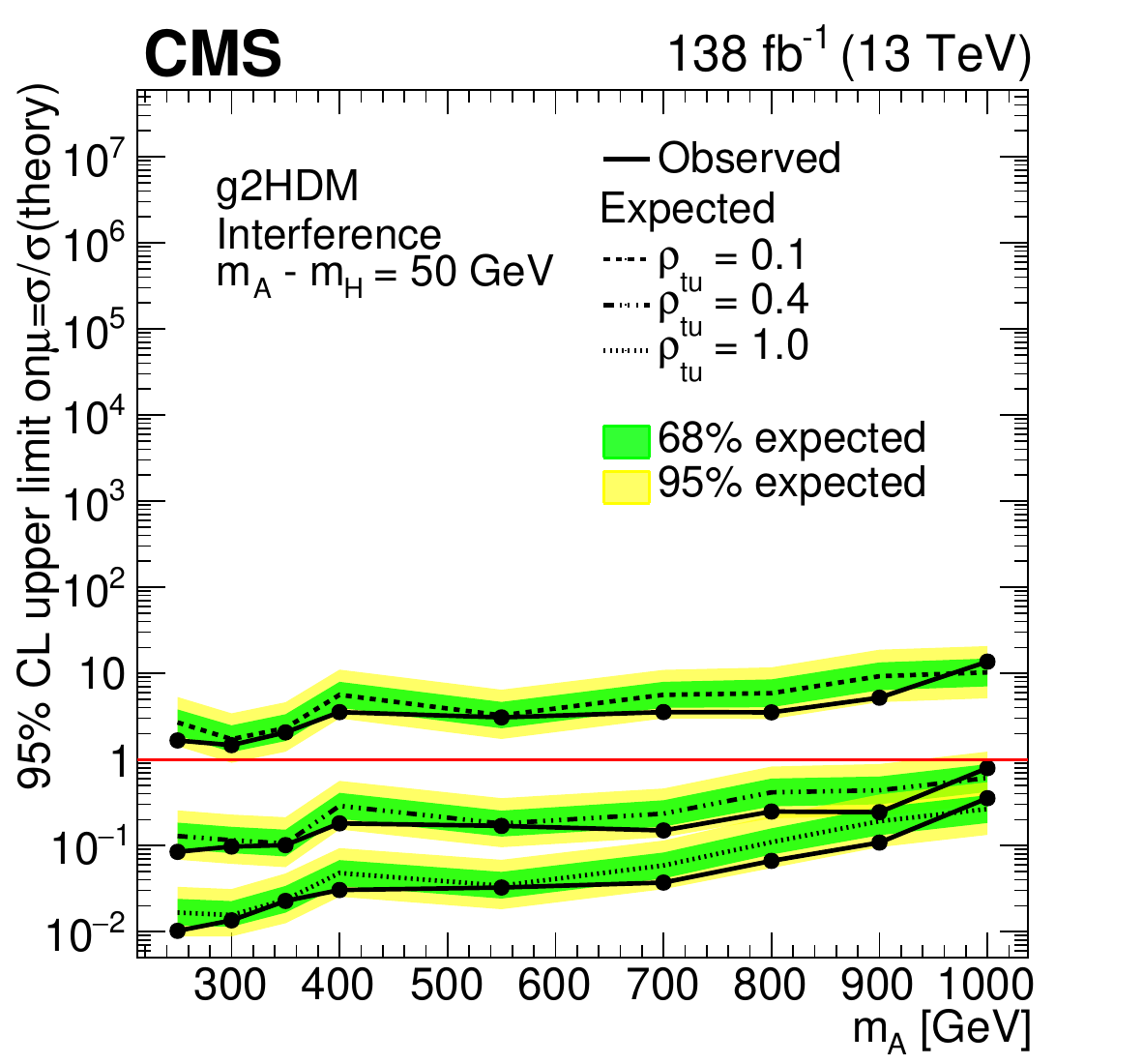}
  \includegraphics[width=0.49\textwidth]{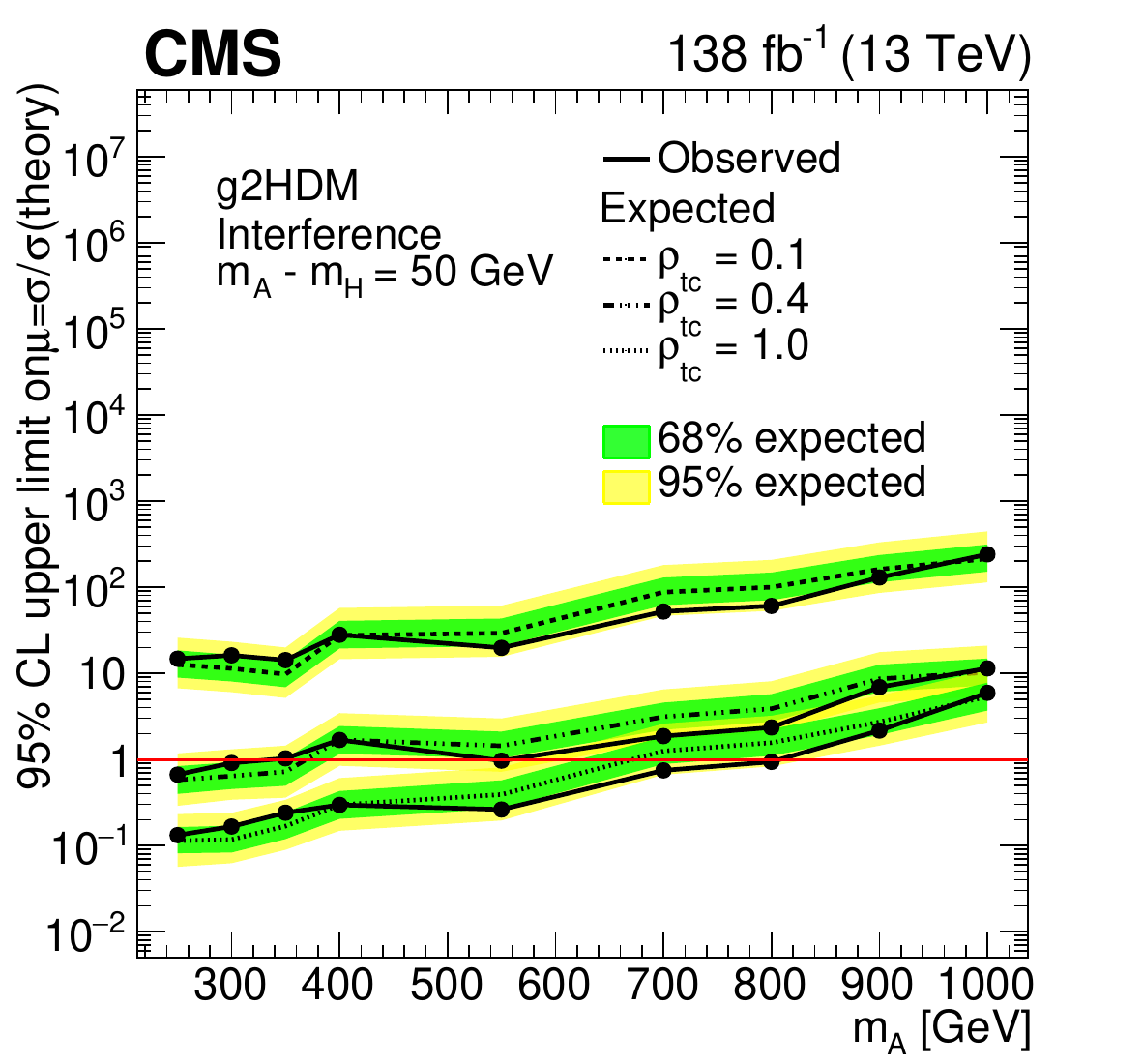}
  \caption{Observed and expected 95\% \CL upper limits on the signal strength as functions of \mA for g2HDM using different coupling assumptions: $\rho_{\PQt\PQu}$ = 0.1, 0.4, 1.0 (\cmsLeft)  and $\rho_{\PQt\PQc}$ = 0.1, 0.4, 1.0 (\cmsRight) with \PA-\PH interference assuming $\mA - \mH = 50\GeV$, for the combination of the \ee, \mumu, and \emu categories and all data-taking periods. The inner (green) band and the outer (yellow) band indicate the regions containing 68 and 95\%, respectively, of the distribution of limits expected under the background-only hypothesis.}
  \label{fig:limit_run2_rtu_rtc_w_interference}
\end{figure}

\begin{figure}[!ht]
  \centering
  \includegraphics[width=0.49\textwidth]{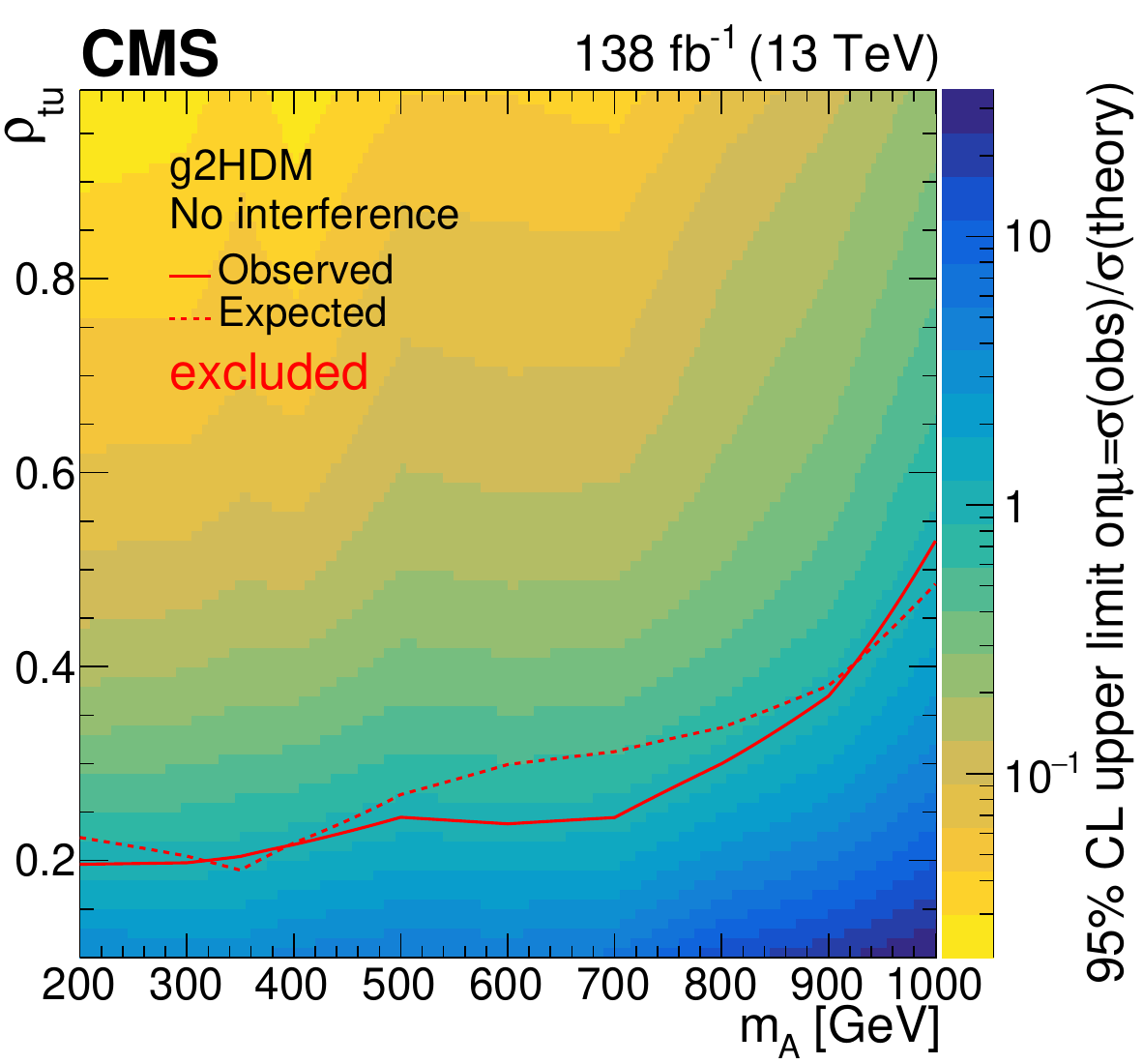}
  \includegraphics[width=0.49\textwidth]{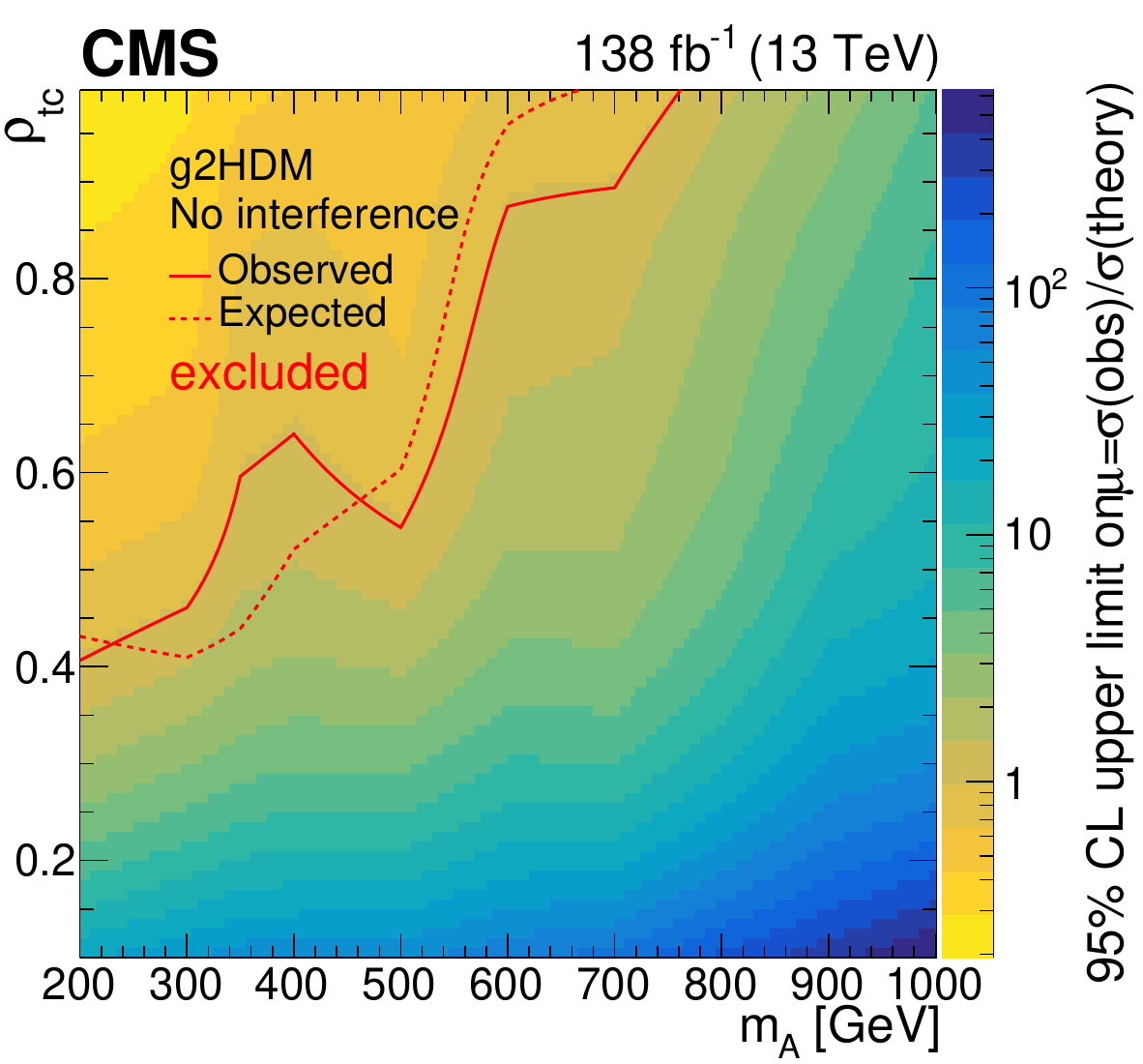}
  \caption{Observed 95\% \CL upper limit on the signal strength as a function of \mA and $\rho_{\PQt\PQu}$ (\cmsLeft) and $\rho_{\PQt\PQc}$ (\cmsRight)  for g2HDM without the \PA-\PH interference, for the combination of the \ee, \mumu, and \emu categories.
The color axis represents the observed upper limit on the signal strength. Expected (dashed lines) and observed (solid lines) exclusion contours are also shown.}
  \label{fig:2Dlimits_noninter}
\end{figure}

\begin{figure}[!ht]
  \centering
  \includegraphics[width=0.49\textwidth]{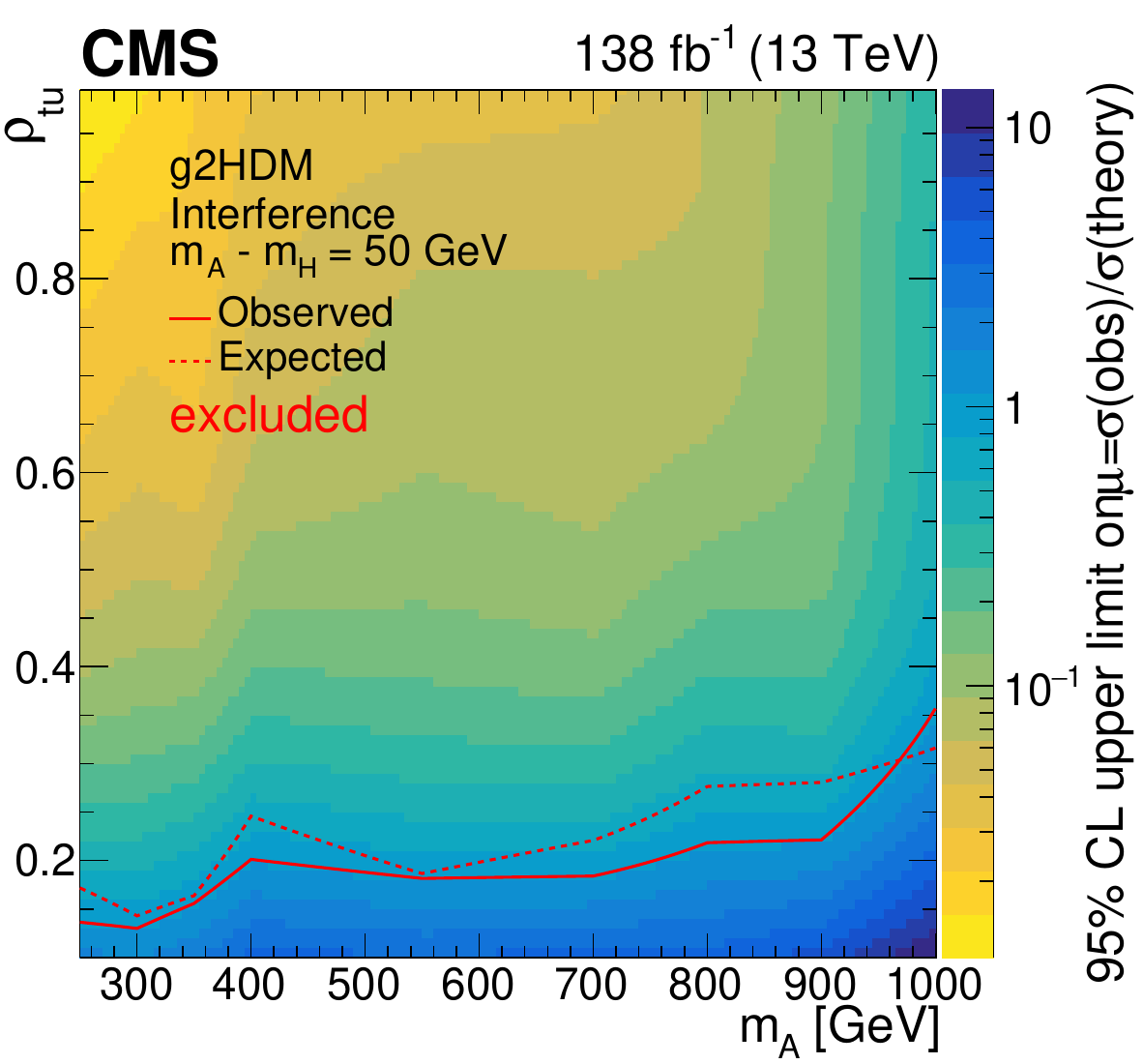}
  \includegraphics[width=0.49\textwidth]{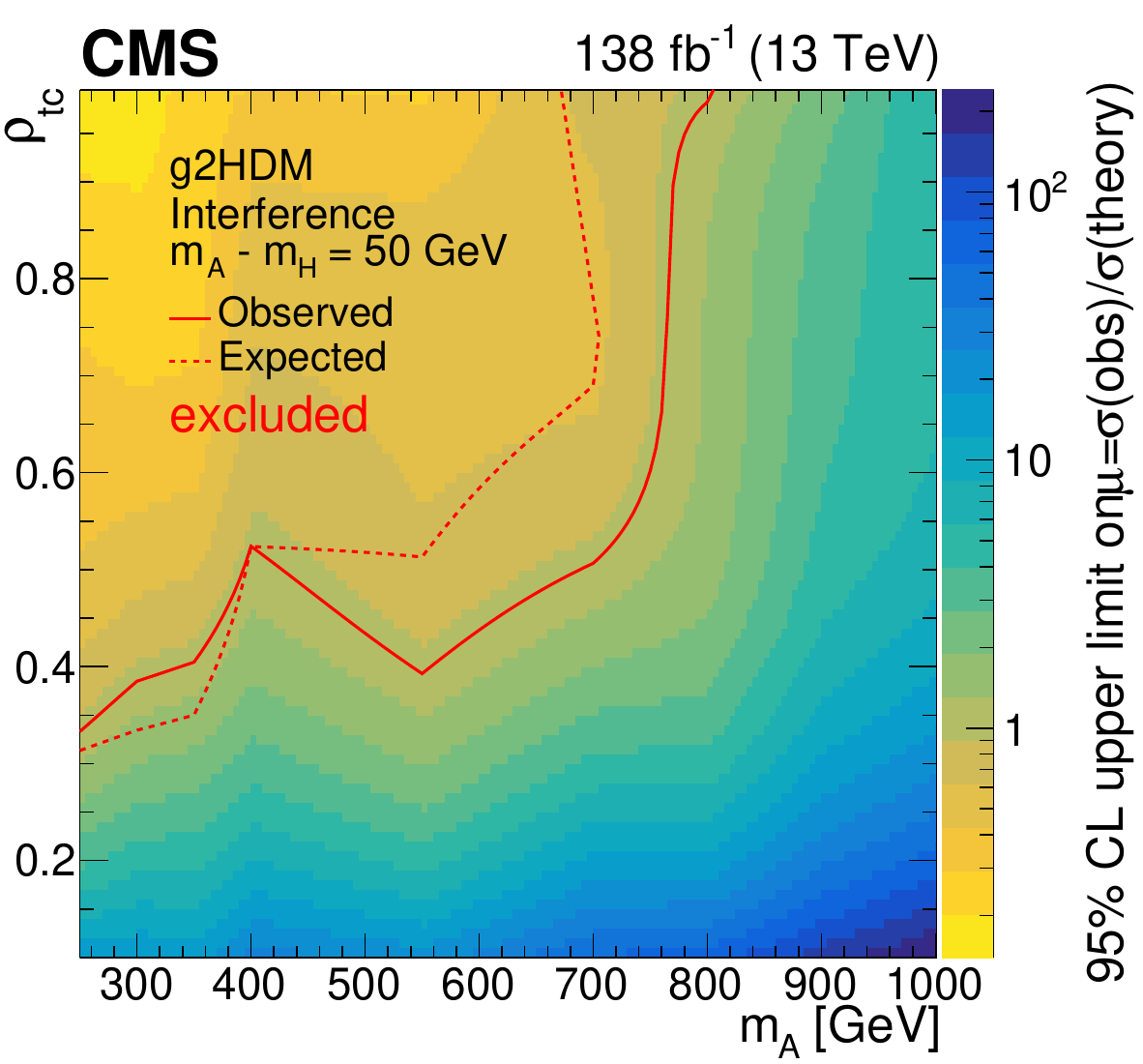}
  \caption{Observed 95\% \CL upper limit on the signal strength as a function of \mA and $\rho_{\PQt\PQu}$ (\cmsLeft) and $\rho_{\PQt\PQc}$ (\cmsRight)  for g2HDM signal model with the \PA-\PH interference, for the combination of the \ee, \mumu, and \emu categories. 
The color axis represents the observed upper limit on the signal strength. Expected (dashed lines) and observed (solid lines) exclusion contours are also shown.}
  \label{fig:2Dlimits_inter}
\end{figure}

No significant excess  over the expected SM background is observed.
The results of the search are interpreted in terms of the upper limits on the signal strength parameter $\mu$ in the g2HDM.
The upper limits on $\mu$ are calculated at 95\% confidence level (\CL) using a modified frequentist method~\cite{Heinemeyer:2013tqa, bib:CLS2, bib:CLS1} computed with an asymptotic approximation~\cite{bib:CLS3}.
The expected 95\% \CL on the production cross section and the regions containing 68 and 95\% of the distribution of the limits expected under the background-only hypothesis are calculated with pseudo-experiments using background-only samples.  
It is observed that the shapes of the BDT distributions are consistent with each other for different coupling values at the same \PA or \PH mass value.
Therefore, for other coupling assumptions than $\rho_{\PQt\PQu/\PQt\PQc}=0.4$, the expected signal yields are scaled separately for $\rho_{\PQt\PQu}$ and $\rho_{\PQt\PQc}$ 
for each mass and coupling hypothesis making use of the corresponding signal cross sections. 
It is also observed that \PA or \PH can be used interchangeably as their corresponding BDT distribution shapes as well as their cross sections are the same. Therefore, we only show the results for \PA in this Letter.  
The observed and expected upper limits are displayed as functions of \mA from 200 to 1000\GeV for $\rho_{\PQt\PQu} = 0.1, 0.4, 1.0$ and $\rho_{\PQt\PQc} = 0.1, 0.4, 1.0$, for the Run-2 data and for the combination of all three decay modes, without (with) \PA-\PH interference in Fig.~\ref{fig:limit_run2_rtu_rtc_wo_interference} (\ref{fig:limit_run2_rtu_rtc_w_interference}).
The observed limits at 95\% \CL on the signal strength as functions of \mA and the new Yukawa couplings $\rho_{\PQt\PQc/\PQt\PQu}$ are shown in Fig.~\ref{fig:2Dlimits_noninter} without \PA-\PH interference, and in Fig.~\ref{fig:2Dlimits_inter} with \PA-\PH interference.
Interpreted in g2HDM, exclusion limits at 95\% \CL on the mass of the additional Higgs bosons \PA and \PH are shown in Table~\ref{tab:limits} for different $\rho_{\PQt\PQu}$ and $\rho_{\PQt\PQc}$ assumptions.  

\begin{table}[!htpb]
\centering
\topcaption{Observed (expected) lower limits on \mA at 95\% \CL. For the scenario without interference, the limits on \mH and \mA are the same.}
 \begin{tabular}{cccc}
	 \multicolumn{4}{c}{Observed (expected) mass limit [\GeVns{}]}\\
               & without       & with         & with \\
               & interference  & interference & interference\\
\hline
               &   \mA or \mH        &     \mA                & \mH \\
$\rho_{\PQt\PQu}$    & & & \\
0.4  &   920 (920) &  1000 (1000) & 950 (950)\\
1.0  &   1000 (1000) & 1000 (1000) & 950 (950)\\
     &                & & \\
$\rho_{\PQt\PQc}$    &     &     &   \\
0.4  &   no limit & 340 (370) & 290 (320)\\
1.0  &   770 (680)& 810 (670) & 760 (620)\\
\end{tabular}
\label{tab:limits}
\end{table}

\section{Summary}
\label{sec:summary}
A search for new Yukawa couplings of the top quark in models with additional Higgs bosons in proton-proton collisions at a center-of-mass energy of 13\TeV has been presented. The process considered is the production of same-sign top quark pairs associated with an up or a charm quark, and resulting in a final state containing two same-sign leptons and jets. No significant excess above the background prediction is observed. When no interference between the pseudoscalar (\PA) and scalar (\PH) Higgs bosons is assumed, \PA or \PH bosons with masses below 920\GeV and 1000\GeV are excluded at the 95\% confidence level (\CL) for coupling values $\rho_{\PQt\PQu} = 0.4$ and 1.0, respectively, while all other extra Yukawa couplings are assumed to be zero. Similarly, without interference between \PH and \PA, and assuming a coupling value of $\rho_{\PQt\PQc} = 1.0$, \PA or \PH bosons with masses below approximately $770\GeV$ are excluded at the 95\% \CL. Under the assumption that \PA and \PH interfere in the scenario with a mass difference of $\mA - \mH = 50\GeV$, the pseudoscalar Higgs boson is excluded for $\mA$ values below $1000\GeV$ when considering coupling values $\rho_{\PQt\PQu} > 0.4$. Furthermore, assuming $\rho_{\PQt\PQc} = 0.4$, the exclusion limit for \PA is $\mA = 340\GeV$, whereas assuming $\rho_{\PQt\PQc} = 1.0$, the limit extends to $\mA = 810\GeV$ at 95\% \CL. These results represent the first search based on the generalized two-Higgs-doublet model considering  \PA-\PH interference. 

\begin{acknowledgments}
We congratulate our colleagues in the CERN accelerator departments for the excellent performance of the LHC and thank the technical and administrative staffs at CERN and at other CMS institutes for their contributions to the success of the CMS effort. In addition, we gratefully acknowledge the computing centers and personnel of the Worldwide LHC Computing Grid and other centers for delivering so effectively the computing infrastructure essential to our analyses. Finally, we acknowledge the enduring support for the construction and operation of the LHC, the CMS detector, and the supporting computing infrastructure provided by the following funding agencies: SC (Armenia), BMBWF and FWF (Austria); FNRS and FWO (Belgium); CNPq, CAPES, FAPERJ, FAPERGS, and FAPESP (Brazil); MES and BNSF (Bulgaria); CERN; CAS, MoST, and NSFC (China); MINCIENCIAS (Colombia); MSES and CSF (Croatia); RIF (Cyprus); SENESCYT (Ecuador); MoER, ERC PUT and ERDF (Estonia); Academy of Finland, MEC, and HIP (Finland); CEA and CNRS/IN2P3 (France); SRNSF (Georgia); BMBF, DFG, and HGF (Germany); GSRI (Greece); NKFIH (Hungary); DAE and DST (India); IPM (Iran); SFI (Ireland); INFN (Italy); MSIP and NRF (Republic of Korea); MES (Latvia); LAS (Lithuania); MOE and UM (Malaysia); BUAP, CINVESTAV, CONACYT, LNS, SEP, and UASLP-FAI (Mexico); MOS (Montenegro); MBIE (New Zealand); PAEC (Pakistan); MES and NSC (Poland); FCT (Portugal); MESTD (Serbia); MCIN/AEI and PCTI (Spain); MOSTR (Sri Lanka); Swiss Funding Agencies (Switzerland); MST (Taipei); MHESI and NSTDA (Thailand); TUBITAK and TENMAK (Turkey); NASU (Ukraine); STFC (United Kingdom); DOE and NSF (USA).

\hyphenation{Rachada-pisek} Individuals have received support from the Marie-Curie program and the European Research Council and Horizon 2020 Grant, contract Nos.\ 675440, 724704, 752730, 758316, 765710, 824093, and COST Action CA16108 (European Union); the Leventis Foundation; the Alfred P.\ Sloan Foundation; the Alexander von Humboldt Foundation; the Science Committee, project no. 22rl-037 (Armenia); the Belgian Federal Science Policy Office; the Fonds pour la Formation \`a la Recherche dans l'Industrie et dans l'Agriculture (FRIA-Belgium); the Agentschap voor Innovatie door Wetenschap en Technologie (IWT-Belgium); the F.R.S.-FNRS and FWO (Belgium) under the ``Excellence of Science -- EOS" -- be.h project n.\ 30820817; the Beijing Municipal Science \& Technology Commission, No. Z191100007219010 and Fundamental Research Funds for the Central Universities (China); the Ministry of Education, Youth and Sports (MEYS) of the Czech Republic; the Shota Rustaveli National Science Foundation, grant FR-22-985 (Georgia); the Deutsche Forschungsgemeinschaft (DFG), under Germany's Excellence Strategy -- EXC 2121 ``Quantum Universe" -- 390833306, and under project number 400140256 - GRK2497; the Hellenic Foundation for Research and Innovation (HFRI), Project Number 2288 (Greece); the Hungarian Academy of Sciences, the New National Excellence Program - \'UNKP, the NKFIH research grants K 124845, K 124850, K 128713, K 128786, K 129058, K 131991, K 133046, K 138136, K 143460, K 143477, 2020-2.2.1-ED-2021-00181, and TKP2021-NKTA-64 (Hungary); the Council of Science and Industrial Research, India; ICSC -- National Research Center for High Performance Computing, Big Data and Quantum Computing, funded by the EU NexGeneration program (Italy); the Latvian Council of Science; the Ministry of Education and Science, project no. 2022/WK/14, and the National Science Center, contracts Opus 2021/41/B/ST2/01369 and 2021/43/B/ST2/01552 (Poland); the Funda\c{c}\~ao para a Ci\^encia e a Tecnologia, grant CEECIND/01334/2018 (Portugal); the National Priorities Research Program by Qatar National Research Fund; MCIN/AEI/10.13039/501100011033, ERDF ``a way of making Europe", and the Programa Estatal de Fomento de la Investigaci{\'o}n Cient{\'i}fica y T{\'e}cnica de Excelencia Mar\'{\i}a de Maeztu, grant MDM-2017-0765 and Programa Severo Ochoa del Principado de Asturias (Spain); the Chulalongkorn Academic into Its 2nd Century Project Advancement Project, and the National Science, Research and Innovation Fund via the Program Management Unit for Human Resources \& Institutional Development, Research and Innovation, grant B05F650021 (Thailand); the Kavli Foundation; the Nvidia Corporation; the SuperMicro Corporation; the Welch Foundation, contract C-1845; and the Weston Havens Foundation (USA).
\end{acknowledgments}

\bibliography{auto_generated}
\cleardoublepage \appendix\section{The CMS Collaboration \label{app:collab}}\begin{sloppypar}\hyphenpenalty=5000\widowpenalty=500\clubpenalty=5000
\cmsinstitute{Yerevan Physics Institute, Yerevan, Armenia}
{\tolerance=6000
A.~Hayrapetyan, A.~Tumasyan\cmsAuthorMark{1}\cmsorcid{0009-0000-0684-6742}
\par}
\cmsinstitute{Institut f\"{u}r Hochenergiephysik, Vienna, Austria}
{\tolerance=6000
W.~Adam\cmsorcid{0000-0001-9099-4341}, J.W.~Andrejkovic, T.~Bergauer\cmsorcid{0000-0002-5786-0293}, S.~Chatterjee\cmsorcid{0000-0003-2660-0349}, K.~Damanakis\cmsorcid{0000-0001-5389-2872}, M.~Dragicevic\cmsorcid{0000-0003-1967-6783}, P.S.~Hussain\cmsorcid{0000-0002-4825-5278}, M.~Jeitler\cmsAuthorMark{2}\cmsorcid{0000-0002-5141-9560}, N.~Krammer\cmsorcid{0000-0002-0548-0985}, A.~Li\cmsorcid{0000-0002-4547-116X}, D.~Liko\cmsorcid{0000-0002-3380-473X}, I.~Mikulec\cmsorcid{0000-0003-0385-2746}, J.~Schieck\cmsAuthorMark{2}\cmsorcid{0000-0002-1058-8093}, R.~Sch\"{o}fbeck\cmsorcid{0000-0002-2332-8784}, D.~Schwarz\cmsorcid{0000-0002-3821-7331}, M.~Sonawane\cmsorcid{0000-0003-0510-7010}, S.~Templ\cmsorcid{0000-0003-3137-5692}, W.~Waltenberger\cmsorcid{0000-0002-6215-7228}, C.-E.~Wulz\cmsAuthorMark{2}\cmsorcid{0000-0001-9226-5812}
% ...
\par}
\vskip\cmsinstskip
\dag:~Deceased\\
$^{1}$Also at Yerevan State University, Yerevan, Armenia\\
$^{2}$Also at TU Wien, Vienna, Austria\\
% ...
\end{sloppypar}
\end{document}